\documentstyle[twocolumn,aps,graphicx]{revtex}

\begin{document}

\title{Localized Random Lasing Modes and a New Path for Observing 
Localization}

\author{Xunya Jiang and C.~M.~Soukoulis}

\address{ Ames Laboratory-USDOE and Department of Physics and Astronomy,\\
Iowa State University, Ames, IA 50011\\}

\author{\parbox[t]{5.5in}
{\small
We demonstrate that a knowledge of the density-of-states and  the 
eigenstates of a random system without gain, in conjunction with the 
frequency profile of the gain, can accurately predict the mode that will 
lase first. Its critical pumping rate can be also obtained. It is found 
that the shape of the wavefunction of the random system remains 
unchanged as gain is introduced. 
These results were obtained by the time-independent transfer matrix 
method and finite-difference-time-domain (FDTD) methods. They can be 
also analytically understood by 
generalizing the semi-classical Lamb theory of lasing in random systems.
These findings provide a new path for observing the localization of light, 
such as looking for mobility edge and studying the localized states. 
 \\ 
PACS~numbers:~42.25.Bs, 72.55.Jr, 72.15.Rn, 05.40.-a\\}} 
\maketitle


Localization theory and laser theory were both developed in the sixties. 
The propagation of quantum and classical waves in disordered media has 
been well understood \cite{local}, while
the
laser physics has been well established \cite{lamb,lasers} at the same time.
It was always assumed that disorder was detrimental to 
lasing action. However, Letokhov \cite{letok} theoretically predicted the 
possibility of lasing in a random system, called 
``random laser". But only after the experimental observations by Lawandy 
{\it et al.} \cite{lawandy}, the random laser systems
have been  studied intensively  \cite
{sha,wiersma,expt1,expt2,john,gena,jiang3,jiangbook,jiang4,jiang,jiang2,pr,been,qiming}.
Since then, many experiments were carried out that showed a drastic 
spectral  narrowing \cite{sha} and a narrowing of the coherent 
backscattering 
peak \cite{wiersma}. Recently, new experiments \cite{expt1,expt2} showed 
random 
laser action with sharp lasing peaks. To fully explain such an unusual 
behavior of stimulated emission in random systems, many
theoretical models were constructed. John and Pang \cite{john} studied the 
random lasing system by combining the electron number equations of energy 
levels with the diffusion equation. Berger {\it et al.} \cite{gena} 
obtained the spectral and spatial evolution of emission from random 
lasers by using a Monte Carlo simulation. Very recently, Jiang and 
Soukoulis \cite{jiang3,jiangbook}, by combining a FDTD method with  
laser theory \cite{lamb,lasers}, were able to study the 
interplay of localization and amplification. They obtained 
\cite{jiang3,jiangbook} the field pattern and the spectral peaks of 
localized lasing 
modes inside the system. They were able to explain \cite{jiang3,jiangbook} 
the multi-peaks and the non-isotropic properties in the emission spectra, 
seen experimentally \cite{expt1,expt2}. 
Finally, the mode repulsion property which gives a saturation of the number 
of lasing
modes in a given random laser system  was predicted.  This prediction was 
checked experimentally by Cao {\it et al.} \cite{jiang4}.

One very interesting question that has not been answered by previous 
studies is what is the form of the wavefunction in a random laser 
system. How does the wavefunction in a random system change as one 
introduces gain? 
Does the wavefunction retain its shape in the presence of gain? 
Another very interesting point is if we can predict 
a priori, which mode will lase first? 
What will be its emission wavelength? 
If we understand these issues, we will be able to 
design random lasers with the desired emission wavelengths. In 
addition, we will be able to use the random laser as a new tool to  
study the localization properties of random systems.       

In this paper, we explore the evolution of the wavefunction without and 
with gain, by the time-independent transfer matrix theory 
\cite{jiang,jiang2,pr,been,qiming},
as well as by time-dependent theory \cite{jiang3,jiangbook}. The emission 
spectra 
can be also obtained. In addition, we can also use the semi-classical 
theory of lasing \cite{lamb,lasers} to obtain analytical results for the 
threshold 
of lasing, as well as which mode will lase first. This depends on 
the gain profile, as well as the quality factor 
$Q$ of the modes before  gain is introduced.

Our system is essentially a one-dimensional simplification of the real 
experiments \cite{expt1,expt2}. It consists of many dielectric layers of 
real 
dielectric constant $\varepsilon_2=2.56 \varepsilon_0$ ($\varepsilon_0$ is 
the dielectric permeability of free space) of fixed thickness ($b_0=100$ 
nm), sandwiched between two surfaces, with the spacing between the 
dielectric layer  assumed to be a random variable $a_n=a_0(1+W)$, 
where  $a_0=200$ nm and $W$ has a random value in the range of 
[-0.7, 0.7]. We choose a 30-cell random system, as the first system of 
our numerical study. In Fig. 1a, we present the results of the logarithm 
of the transmission coefficient as a function of frequency $f$. These 
results were obtained by using the transfer matrix techniques introduced 
in Ref. 15. Notice that we have three typical resonance peaks (denoted 
$P_1$, $P_2$ and $P_3$) in the frequency range of 600 to 660 THz. As 
one can see from Fig. 1a, the linewidths of the three modes are different. 
$P_3$ has the smallest linewidth and therefore the largest $Q$, while 
$P_1$ has the largest linewidth and therefore the smallest $Q$. We have 
also numerically calculated the wavefunctions corresponding to these three 
peaks and indeed find out that the more localized wavefunction is the one 
with the larger $Q$. All the results above were obtained for the case 
without gain. 

According to the semiclassical theory of laser physics 
\cite{lamb,lasers}, we generally use a polarization due to gain 
$P_{gain}=\varepsilon_0\chi(\omega)E
=\varepsilon_0(\chi'(\omega)+i\chi''(\omega))E $
to introduce amplifying medium effects. 
Both 
$\chi'(\omega)$ and $\chi''(\omega)$ are proportional to the outside pumping 
rate $P_r$ and can be expressed by the parameters of the gain material
\cite{gain}.  

To determine which peak will lase first, we can again use the 
time-independent transfer matrix method (see Eq. (4) of Ref. 15) with a 
frequency-$independent$ gain, which means the width of the gain profile 
is very large. 
It is well understood that time-independent 
theory \cite{jiang,jiang2,pr,been,qiming} for random lasers can be used to 
obtain the threshold 
for lasing. At threshold, the transmission  
coefficient $T$ goes to infinity.  In Fig. 1b, we plot the $log_{10}(T)$ 
versus $\chi''$ 
for the three peaks $P_1$, $P_2$ and $P_3$. 
In Fig. 1b,  the 
$P_3$ which has the largest $Q$ has the smallest threshold for lasing. 
So, in the 
frequency-independent gain case, the 
transfer matrix 
method gives that the mode with largest  $Q$ will lase first. We have 
also used the transfer-matrix method with a frequency-dependent gain 
profile, given as a dotted line in Fig. 1a. We choose the central 
frequency as $f_a=\omega_a/2\pi=618.56$ THz, which is exactly between the 
$P_1$ and $P_2$, 
and its width to be $\Delta f_a=\Delta\omega_a/2\pi=15$ THz. In this 
case, when we increase the pumping rate $P_r$, we find the $P_2$ 
will lase first, then $P_1$. So two important conditions  
determine which mode will lase first, the first is the quality factor of 
the mode and the second is the gain profile.     
Experimentally, it is quite often that only part of the random medium has 
been pumped. Then a third factor comes in, i.e. the spatial overlap of 
the mode function and the gain region.

\begin{figure*}[h]
\begin{center}
\includegraphics[height=9.cm,width=6.cm,angle=270]{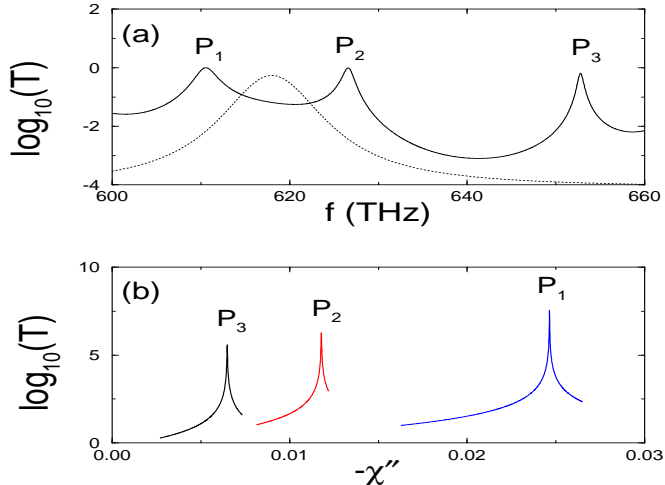}
\caption{(a) Logarithm of transmission coefficient vs frequency of a 30-cell
random system with three typical peaks. The dotted line
shows the frequency dependent of the gain profile. (b) Logarithm of the
transmission coefficient vs  $\chi''$ for the three peaks shown
in Fig. 1a with a frequency-independent gain. $\chi''$ is 
proportional to the pumping rate $P_r$.} 
\end{center}
\end{figure*}

The next issue we address is the shape of the wave function, as one 
introduces gain. In Fig. 2a, we present the amplitude of the electric 
field versus distance for $f=626.6$ THz, which corresponds to the peak 
$P_2$ of Fig. 1a. 
 In Fig. 2b, 
we show  the wavefunction 
with near-threshold gain 
at the exact high-peak 
frequency of 
$P_2$. Both Fig. 2a and 2b are obtained by the time-independent 
transfer matrix method. 
Notice that in 
the presence of gain, the shape of the wavefunction in Fig. 2b is almost 
same as that 
without gain in Fig. 2a. The only change is its amplitude which increases 
uniformly. Actually, by keeping the incident amplitude the same, 
when we increase the gain
from zero to the threshold value, we find that the 
amplitude of the wavefunction  
increases from a small to a very large value, but its shape remains almost 
same. 

\begin{figure}[h]
\includegraphics[height=8cm,width=8cm,angle=270]{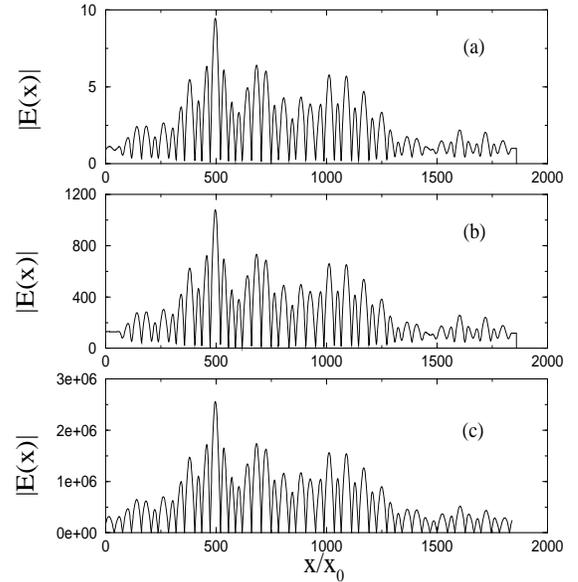}
\caption{Amplitude of the electric field of the 30-cell system vs
$x/x_0$, with $x_0=5$ nm, for the peak $P_2$ in Fig. 1a. (a) Without gain,
incident field $E_{inc}=1 $V/m; (b) With near-threshold gain ( $P_r=7.2
\times 10^6 $/s ), incident
field $E_{inc}=1 $V/m; (c) The lasing field with over-threshold pumping
rate ($P_r=2 \times 10^7 $/s . Both (a) and (b) are obtained by the transfer
matrix method while 
(c) is obtained by the FDTD method and 
laser theory. } 
\end{figure}

To check if indeed 
this amazing property is also obeyed by the full time-dependent 
theory \cite{jiang3,jiangbook} of the semiclassical laser theories with 
Maxwell-equations, we repeat our theory for this case too. While 
in the time-independent theory, every mode is independent and 
amplification is not saturated, this is not true for a real lasing 
system. In a real lasing system, modes will compete with each 
other. As was discussed in Ref. 12, in a $short$ system, the first 
lasing mode will suppress all other modes, so we observe only one lasing 
mode even for a large pumping rate. This is exactly the case when 
we use the FDTD method to simulate  
our 30-cell system with an over-threshold gain whose gain profile is the
same as given in Fig.1a. At first, the electric field is  a random one due 
to   
spontaneous emission, then a strong lasing mode evolves from the noisy 
background and a sharp peak appears in the emission spectrum after 
 Fourier transform. The stable field profile of the 
FDTD calculation is given in Fig. 2c, where the 
wavefunction
is quite the same as in Fig. 2a and 2b, but the amplitude is very 
large. The emission spectrum of this case is obtained, and  
gives a sharp peak very close to the resonant peak $P_2$ of Fig. 1a. 
We also checked the wavefunctions as well as the lasing 
threshold as we shift the gain profile. If the central 
frequency of the gain profile is near $P_3$, the $P_3$ mode will lase 
first, and the shape of wavefunction of $P_3$ remains unchanged.
We have also checked the above ideas for a larger ($L\gg\xi$, where 
$\xi$ is the localization length) system.
In such strong localized case, the form of the wavefunction remains 
unchanged as one introduces gain.



The numerical results which are presented above  can be explained by 
the 
semi-classical Lamb theory \cite{lamb} of laser physics. According to Lamb 
theory,
the Maxwell equation of the random laser system can be written as

\begin{eqnarray}
 -{\nabla}^2  {E(x,t) } + \mu_0 \sigma \frac{\partial E(x,t)}{\partial
t} +  \mu_0 \varepsilon(x) \frac{\partial^2 E(x,t)}{\partial t^2 }  = -\mu_0
\frac{\partial^2 P_{gain}(x,t)}{\partial t^2 }
\end{eqnarray}
Where $\mu_0$ is the permeability of free space, $E(x,t)$ is the electric 
field 
, and the dielectric constant $\varepsilon(x)$
is determined by the random configuration of the system. $\sigma$ is
not just the common conductivity loss, but can be interpreted by the  
total mode loss including the    surface loss of the system   
by radiation. The polarization due to gain $P_{gain}(x,t)$ \cite{gain} is 
the same as the one defined above.
 
We assume 
that the system satisfies the slow varying approximation
(it is always satisfied if we care about the stable lasing state). So
$E(x,t)=E_m(t)\phi_m(x)exp(-i \omega t)$ (our later discussion shows that 
the separation of spatial and time parts of wavefunction is reasonable), 
where $E_m(t)$ is the field amplitude, $\phi_m(x)$ is the normalized 
wavefunction, and 
$\omega $ is the frequency of the field. The surface loss of the mode is
$\sigma_m =\varepsilon_0 \omega/Q_m$, where the $Q_m$ is the quality
factor of the mode. Then, we  get two equations \cite{lamb} for the real 
and the imaginary terms of Eq. (1)

\begin{eqnarray}
 {\nabla}^2  {\phi_m(x) }
+  \mu_0 (\varepsilon(x)+\varepsilon_0 \chi'(x, \omega)) \omega^2 \phi_m(x) 
= 0
\end{eqnarray}

\begin{eqnarray}
\frac{\partial E_m(t) }{\partial t} =
 \left( -\overline {\chi''(\omega)}
-\frac {1}{Q_m } \right) \frac{\varepsilon_0\omega E_m(t)}{2 
\overline{\varepsilon}} 
\end{eqnarray}
where $\overline{\varepsilon}=\int_0^L\varepsilon(x) dx/L$ is the 
spatial averaged dielectric constant inside the system, and 
$\overline {\chi''(\omega)}=\int_0^L \phi_m(x)^\ast\chi''(x, 
\omega)\phi_m(x)dx/L $ is the spatial averaged gain. The last integral is 
done to take into account of  the overlap between the wavefunction and  the 
spatial region of gain. 

Equation (2)  determines the field distribution, quality factor  $Q_m$, 
and the vibration frequency $\omega$ of the lasing mode.
The term $\varepsilon_0 \chi'(x, \omega)$ will cause the vibration frequency
to shift away from the original eigenfrequency of the mode, called 
{\it pulling effect}. For a well-defined mode, generally
$Q_m \gg 1$,  we need a very small gain to lase. Then $\chi'(x, \omega) \ll
1 $,  the pulling effect is very weak, and $\omega \simeq \Omega_m$, where
$\Omega_m$ is the eigenfrequency of the mode. So the wavefunction of the
lasing mode  should be similar to the eigenfunction of the mode.  
Theoretically, we can use the perturbation method to obtain $\omega$ and the 
wavefunction.

Equation (3) is the time-dependent amplitude equation.  First we can use it 
to  determines 
 the threshold condition when $ -\overline  \chi''(\omega)={1}/{Q_m }$. 
For our system, with homogeneous pumping, we have 
\begin{eqnarray}
P_r^c =C_0  \frac {1+ \frac{4(\omega_a-\Omega_m)^2}{\Delta\omega_a^2} }
{Q_m}
\end{eqnarray}
where
$C_0=\frac{a_0+b_0}{a_0}\frac{\gamma_c m_e \varepsilon_0 \omega_a \Delta
\omega_a} {\gamma_r N_0\tau_{21}e^2}$
is a constant. 
Eq. (4) indeed gives that the threshold value of lasing is inversely
proportional to the quality factor $Q$.

Second, Eq. (3) give us the stable amplitude of field when the gain is 
over threshold.  
Actually, the gain is saturable, $\chi''(\omega) \propto \Delta N  
\propto 1/(1+C_2|E_m|^2)$ \cite{gain}, in real systems and in our 
FDTD model 
\cite{jiang3,jiangbook}.  With an over-threshold gain,  $E_m(t)$ will 
increase and the gain parameter $\chi''(\omega, E_m)$ will decrease 
until $-\overline\chi''(\omega, E_m)=1/Q_m$, then the field is stable. 
So Eq. (3) also 
determines the amplitude of the stable field for over-threshold
pumping cases.

\begin{figure}[h]
\includegraphics[height=8cm,width=8cm,angle=270]{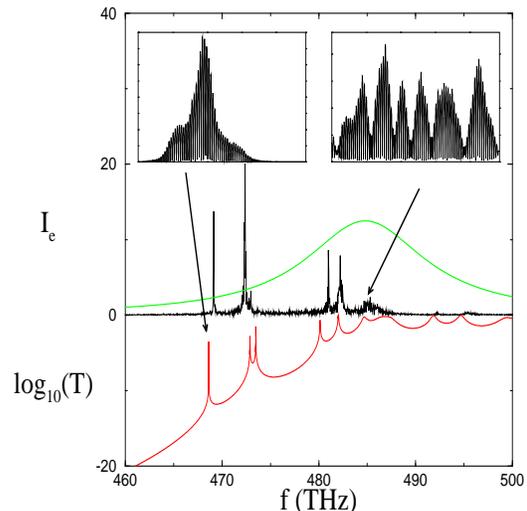}
\caption{
Logarithm of the transmission coefficient $T$, the emission intensity   
$I_e$ and the gain profile vs frequency $f$ for a 80-cell random systems.
In the insert, wavefunctions of two modes are shown. }
\end{figure}

Our numerical and analytical results clearly suggest that
states of the random system with gain can easily lase, provided
their $Q$-factor is large.  These findings provide a new path for observing
localization of light. 
  Since localized states have large $Q$ values, 
will lase with a  small pumping rate. On the other hand, strongly
fluctuating extended states have smaller $Q$ values because of the 
radiation loss on the surface of the system and can lase only 
after a stronger pumping. In a real experiment, even in the presence of
absorption, if the gain profile is close to the mobility edge, there is
going to be a discontinuity in the critical pumping rate needed for lasing.
Localized states will lase first at a low pumping rate,
while extended states need a high pumping rate. 
In Fig. 3, we present the results of the DOS vs frequency $f$ for a 1d 
quasi-disordered system. On the insert two eigenfunctions are given, one 
localized at $f=468$ THz and the one ``extended" at  $f=485$ THz. In 
addition the emission intensity $I_e$ vs $f$ is also shown with gain, 
$P_r=10^5$ $1/s$, which demonstrate that the localized modes will lase 
first. In a more realistic 3D case, we expect $I_e$ vs frequency to have 
many peaks for the localized region and no peaks in the extended region 
for a give pumping rate \cite{skhunor}. 
It would be very interesting if this discontinuity can be observed 
experimentally.



In summary, we have used the time-independent transfer matrix
method and the FDTD method to show that all lasing modes are coming from
the eigenstates of the random system. A knowledge of the eigenstates and
the density-of-states of the random system, in conjunction with the
frequency profile of the gain, can accurately predict the mode that will
lase first, as well as its critical pumping rate. Our detailed numerical
results clearly demonstrate that the shape of the wavefunction remains
unchanged as gain is introduced into the system. The role of the gain is to
just increase uniformly the amplitude of the wavefunction without
changing its shape. These results can be understood by generalizing the
semi-classical Lamb theory of lasing in random systems. These findings
can help us unravel the conditions of observing the localization of light, as
well as manufacturing random laser with specific emission wavelengths.

Ames Laboratory is operated for the U.S. Department of Energy by Iowa
State University under Contract No. W-7405-Eng-82. This work was
supported by the director for Energy Research, Office of Basic Energy
Sciences.





\begin{thebibliography}{s1}





\bibitem{local}For a recent review see C.~M.~Soukoulis and 
E.~N.~Economou, Waves Random Media {\bf 9}, 255 (1999).


\bibitem{lamb}
A.~Maitland and M.~H.~Dunn, {\it Laser Physics} (North-Holland Publishing
Com., Amsterdam, 1969), chapter 9. 


\bibitem{lasers} Anthony~E.~Siegman, {\it Lasers} (Mill Valley,
California, 1986), chapters 2, 3, 6 and 13.



\bibitem{letok}V.~S.~Letokhov, Sov. Phys. JEPT {\bf 26}, 835 (1968).


\bibitem{lawandy}N.~M.~Lawandy, R.~M.~Balachandran, S.~S.~Gomers, and
E.~Sauvain, Nature {\bf 368}, 436 (1994).

\bibitem{sha}W.~L.~Sha, C.~H.~Liu and R.~R.~Alfano, Opt. Lett. {\bf 19}, 
1922 (1994); R.~M.~Balachandran and N.~M.~Lawandy, Opt. Lett. {\bf 20}, 
1271 (1995); M.~Zhang, N.~Cue and K.~M.~Yoo, Opt. Lett. {\bf 20}, 961 (1995);
G.~Van~Soest, M.~Tomita and A. Lagendijk, Opt. Lett. {\bf 24}, 306 
(1999); G. Zacharakis {\it et. al.} Opt. Lett. {\bf 25}, 923 (2000). 

\bibitem{wiersma}D.~S.~Wiersma, M.~P.~van
Albada and Ad Lagendijk, Phys. Rev. Lett. {\bf 75}, 1739 (1995);
D.~S.~Wiersma and A.~Lagendijk, Phys. Rev. E {\bf 54}, 4256(1996).


\bibitem{expt1}H.~Cao {\it et al.}
Phys. Rev. Lett. {\bf 82}, 2278 (1999); H.~Cao {\it et al.}
Phys. Rev. Lett. {\bf 84} 5584 (2000). 

\bibitem{expt2}S.~V.~Frolov,
Z.~V.~Vardeny, K.~Yoshino, A.~Zakhidov and R.~H.~Baughman, Phys. Rev. B
{\bf 59}, R5284 (1999).




\bibitem{john}
S.~John and G.~Pang, Phys. Rev. A {\bf 54}, 3642 (1996), and references
therein.

\bibitem{gena}G. A. Berger, M. Kempe, and A. Z. Genack, {\it Phys. Rev. E}
{\bf 56}, 6118 (1997).

\bibitem{jiang3}Xunya~Jiang and C.~M.~Soukoulis, Phys. Rev. Lett. {\bf
85}
70 (2000).

\bibitem{jiangbook}Xunya~Jiang and C.~M.~Soukoulis, in {\it Photonic 
Crystals and Light Localization in the 21st Century} edited by 
C.~M.~Soukoulis (Kluwer, Dordrecht, 2001), p. 417.

  
\bibitem{jiang4}H.~Cao, Xunya Jiang, C.~M~.Soukoulis {\it et al.} 
unpublished.

\bibitem{jiang}Xunya~Jiang and C.~M.~Soukoulis,  Phys. Rev. B {\bf 59},
6159 (1999).


\bibitem{jiang2}Xunya~Jiang, Qiming Li and C.~M.~Soukoulis, Phys. Rev. B
{\bf 59} R9007 (1999).



\bibitem{pr}P.~Pradhan, N.~Kumar, Phys. Rev. B {\bf 50}, 9644 
(1994); Z.~Q.~Zhang, Phys. Rev. B {\bf 52}, 7960 (1995).


\bibitem{been}C.~W.~J.~Beenakker Phys. Rev. Lett. {\bf 81}, 1829 (1998)



\bibitem{qiming}Qiming~Li, K.~M.~Ho and C.~M.~Soukoulis, Physica B {\bf 
296}, 78 (2001).

  




\bibitem{gain}$\chi'(\omega)=\chi''_0 \frac{-\Delta x}{1+\Delta x^2}$ and  
$\chi''(\omega)=\chi''_0\frac{1}{1+\Delta x^2} $, where 
$\chi''_0=\frac{\gamma_r}{\gamma_c}\frac{\Delta N e^2}{\varepsilon_0 
m_e \omega_a  \Delta \omega_a}$ 
and $\Delta x=2 \frac{\omega - \omega_a} {\Delta \omega_a}$,
$\tau_{21}=1/\gamma_r=1\times10^{-10}$ s  is the real lifetime of  
upper lasing level, 
$\gamma_c=\frac{e^2}{m}\frac{{\omega_a}^2}{6\pi\varepsilon_0c^3}$ is the
classical decay rate,   
$\Delta 
N=N_2-N_1$ is number inversion of electrons at upper  and 
lower lasing levels, actually $\chi''(\omega, E_m) \propto \Delta N 
\propto 1/( 1 + C_2 |E_m|^2) $ in our FDTD model and real experimental 
systems so that the gain is saturable for large field amplitude, but when 
the field is weak ($< 10^5$ V/m),  
 the factor $1/( 1 + C_2 |E_m|^2)$ is very close to one; $\omega_a$ and 
$\Delta \omega_a$ are the central frequency 
and linewidth of the gain profile; $e$ and $m_e$ are the charge and mass of 
the electron. For a four-level gain medium, 
$\Delta N = P_r N_0 \tau_{21}$,     
where $N_0=3.01 \times 10^{25}$$1/m^3$ is the electron density of gain 
medium, and $P_r$ is the pumping rate. With our parameters, we have
$P_r= -1.003\times 10^{-21}\chi''_0 \omega_a \Delta \omega_a$ .
For details see Chap. 2 and 3 and 7 of Ref. 3. 

\bibitem{skhunor} see Fig. 7c in M.~N.~Skhunor et. al, Synthetic Metals, 
{\bf 116} 485 (2001)



\end{thebibliography}
\end{document}